\documentclass[preprint,12pt]{elsarticle}

\usepackage{amssymb}

\usepackage{url}

\usepackage{amsmath}

\usepackage{color}
\usepackage{longtable}
\usepackage{xtab}

\journal{Aerospace Science And Technology}

\begin{document}

\begin{frontmatter}



\title{TI tether rig for solving secular spinrate change problem of
  electric sail}


\author[FMI]{Pekka Janhunen\corref{cor1}}
\ead{pekka.janhunen@fmi.fi}
\ead[url]{http://www.electric-sailing.fi}
\author[FMI]{Petri Toivanen}

\address[FMI]{Finnish Meteorological Institute, Helsinki, Finland}
\cortext[cor1]{Corresponding author}

\begin{abstract}
The electric solar wind sail (E-sail) is a way to propel a spacecraft
by using the natural solar wind as a thrust source. The problem of
secular spinrate change was identified earlier which is due to the
orbital Coriolis effect and tends to slowly increase or decrease the
sail's spinrate, depending on which way the sail is inclined with
respect to the solar wind.  Here we present an E-sail design and its
associated control algorithm which enable spinrate control during
propulsive flight by the E-sail effect itself. In the design, every
other maintether (``T-tether'') is galvanically connected through the
remote unit with the two adjacent auxtethers, while the other
maintethers (``I-tethers'') are insulated from the tethers. This
enables one to effectively control the maintether and auxtether
voltages separately, which in turn enables spinrate control.  We use a
detailed numerical simulation to show that the algorithm can fully
control the E-sail's spin state in real solar wind.  The simulation
includes a simple and realistic set of controller sensors: an imager
to detect remote unit angular positions and a vector
accelerometer. The imager resolution requirement is modest and the
accelerometer noise requirement is feasible to achieve. The
TI tether rig enables building E-sails that are able to control their
spin state fully and yet are actuated by pure tether voltage
modulation from the main spacecraft and requiring no functionalities
from the remote units during flight.
\end{abstract}


\end{frontmatter}



\section*{Nomenclature}
\nobreak\noindent
\begin{longtable}{ll}
au     & Astronomical unit, 149\,597\,871\,km \\
$A$    & Auxiliary factor \\
$\mathrm{clamp}\,(x,a,b)$ & Clamp function, limitation of $x$ in $[a,b]$ \\
$d_\mathrm{max}$ & Maximum thrust reduction for $f_4$, 0.05 \\
$dF/dz$ & Thrust per unit length produced by tether \\
$\hat{\mathbf{e}}_r$ & Radial unit vector \\
$f(t)$  & Generic function of time $t$ \\
$f_1(t),f_2(t),\tilde{f}(t)$ & Gap filler functions\\
$f$     & Total throttling factor \\
$f_1,f_2,f_3$ & Individual throttling factors \\
$f_4,f_5$ & Throttling factors for oscillation damping \\
$f_6$ & Throttling factor for setting thrust \\
$f_6^\mathrm{max}$ & Maximum allowed $f_6$, 1.01 \\
$f_6^\mathrm{old}$ & Previous value of $f_6$ \\
$\mathbf{F}$ & Generic thrust vector \\
$F_\mathrm{goal}$           & Goal E-sail thrust, 100 mN \\
$\mathbf{F}_n$ & Spinplane normal component of thrust \\
$\mathbf{F}_\mathrm{rig}$   & Thrust on tether rig \\
$\mathbf{F}_s$ & Spinplane component of thrust \\
$\mathbf{F}_\mathrm{sc}$    & Thrust on spacecraft \\
$\mathbf{F}_\mathrm{tot}$   & Total thrust, $\mathbf{F}_\mathrm{sc}+\mathbf{F}_\mathrm{rig}$ \\
$\mathbf{F}_\mathrm{tot}^\mathrm{ave}$ & Time-averaged version of $\mathbf{F}_\mathrm{tot}$ \\
$F_0$  & Typical tether tension \\
$g$ & Acceleration due to gravity, 9.81 m/s$^2$ \\
$g_\mathrm{d}$   & Greediness factor for damping in $f_4$, 3.0 \\
$g_\mathrm{s}$   & Greediness factor for spinrate change, 2.0 \\
$g_\mathrm{t}$   & Greediness factor for spinplane turning, 1.0 \\
$K$    & Spin axis orientation keeper factor \\
$\mathbf{L}$    & Angular momentum vector \\
$\mathbf{L}(0)$  & Initial angular momentum vector \\
$m_p$ & Proton mass \\
$m_\mathrm{rig}$ & Mass of tether rig, 11 kg \\
$m_\mathrm{sc}$ & Mass of spacecraft body, 300 kg \\
$m_\mathrm{tot}$ & Total mass, 311 kg \\
$\max(a,b)$ & Maximum of $a$ and $b$ \\
$\min(a,b)$ & Minimum of $a$ and $b$ \\
$\hat{\mathbf{n}}_\mathrm{goal}$ & Goal orientation unit vector of spin axis \\
$\hat{\mathbf{n}}_\mathrm{SW}$ & Unit vector along (nominal) SW, (0,0,1) \\
$N_\mathrm{w}$ & Number of tethers \\
$\mathbf{p}$ & Momentum of tether rig \\
$P_\mathrm{dyn}^\perp$ & Solar wind dynamic pressure due to tether-perpendicular flow\\
$\mathbf{r}$    & Position of remote unit \\
$\hat{\mathbf{s}}$ & Unit vector along spin axis \\
$S$    & Spinrate increase factor \\
$t$    & Time \\
$t_1$,$t_2$ & Starttime and endtime of data gap \\
$\mathbf{v}$ & Velocity of remote unit \\
$v_\mathrm{s}$   & Spin axis aligned speed of remote units \\
$v_\mathrm{tot}$ & Average rotation speed of remote units \\
$\mathbf{v}_\perp$ & Tether-perpendicular component of solar wind velocity \\
$V_0$ & Tether voltage \\
$V_1$ & Voltage corresponding to solar wind proton kinetic energy\\
$x,y,z$ & Cartesian coordinates in inertial frame \\
$x'$,$y'$,$z'$ & Spin axis aligned Cartesian coordinates \\
$\hat{\mathbf{x}}',\hat{\mathbf{y}}',\hat{\mathbf{z}}'$ & Unit vectors along $x'$, $y'$, $z'$ \\
$\alpha$ & Sail angle, angle between SW and spin axis \\
$\Delta t$ & Timestep how often controller is called, 2 s \\
$\Delta t_\mathrm{d}$ & How often damper is called, 20 s \\
$\epsilon_0$ & Vacuum permittivity \\
$\phi$ & Polar angle of spin axis vector \\
$\rho$ & Solar wind mass density \\
$\tau_\mathrm{d5}$,$\tau_\mathrm{d6}$ & Timescale parameters, 1200 s \\
$\omega$ & Angular frequency of the sail spin \\
$\Omega$ & Angular frequency of heliocentric orbit \\
\end{longtable}\addtocounter{table}{-1} 

\section{Introduction}

\begin{figure}[h]
\centering
\includegraphics[width=\columnwidth]{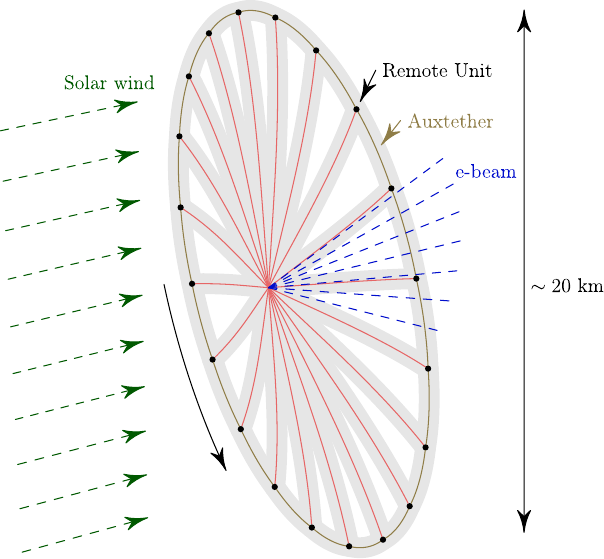}
\caption{
Schematic presentation of the E-sail.
}
\label{fig:Esail}
\end{figure}

The solar wind electric sail (E-sail) is a concept how to propel a
spacecraft in the solar system using the natural solar wind (SW)
\citep{paper1,RSIpaper}. The E-sail uses a number of thin metallic and
centrifugally stretched tethers which are biased at high positive
potential (Fig.~\ref{fig:Esail}). The biasing is effected by an
onboard electron gun which continuously pumps out negative charge from
the tethers.

Thrust vectoring can be done by turning the spin plane by differential
modulation of the tether voltages in sync with the rotation
\citep{paper5}. In this way one can also generate a thrust component
which is perpendicular to the solar wind so that one can e.g.~spiral
outward or inward in the solar system. The thrust magnitude can be
throttled by reducing the voltage and current of the electron
gun. Hence both thrust direction and magnitude can be controlled,
which makes the E-sail a generic method for moving around in the solar
system (outside Earth's magnetosphere) without consuming
propellant. For example, it was demonstrated numerically that one can
reach Mars by the E-sail, even using a simple control law, despite
persistent variations of the solar wind density and the solar wind
flow velocity vector \citep{paper5}.

The following secular spinrate change problem was, however, identified
\citep{paper14}. When an E-sail orbits around the sun with the sail
inclined with respect to the SW, the orbital Coriolis
effect causes a secular increase or decrease of the
spinrate. Inclining the sail is necessary if one wants to produce
transverse thrust perpendicular to the SW direction, which is
usually the case. Specifically, if the sail is inclined so that it
brakes the orbital motion and keeps the spacecraft spiralling towards
the sun, the spinrate decreases, and if the sail is inclined
in the opposite way so that the orbit is an outward moving spiral, the
spinrate increases. The rate of spinrate increase or decrease obeys
approximately the equation \citep{paper14}
\begin{equation}
\omega(t) \approx \omega(0) e^{\pm(\Omega\tan\alpha)t}.
\end{equation}
Here $\Omega$ is the angular frequency of the heliocentric orbit and
$\alpha$ is the sail angle, i.e.~the (positive) angle between the sail spin axis
and the SW direction. For example if $\alpha$ is $35^{\circ}$ and
the spacecraft is in a circular orbit at 1 au distance, the spinrate
changes by 9\,\% in each week. To overcome the problem, various
technical solutions were proposed and analysed, for example the use of
ionic liquid field-effect electric propulsion (FEEP) thrusters
\citep{D462,D463,MarcuccioEtAl2009} or photonic blades \citep{paper16}
on the remote units.

In this paper we present a novel design concept (the TI tether rig)
for the E-sail which overcomes the secular spinrate problem and yields
a technically simple hardware. We also present a control algorithm and
demonstrate by detailed numerical simulation that the algorithm is
able to fly the E-sail in real SW with full capability to
control the orientation of the spin plane and the spinrate. We also
demonstrate that the algorithm is able to accomplish its task using
a simple set of sensors (remote unit position imager and vector
accelerometer) with realistic amount of measurement noise.

The structure of the paper is as follows. We show that electric
auxtethers enable spinrate control, present the TI tether rig design,
the control algorithm, the dynamical simulation model and the
simulation results. The paper closes with summary and conclusions.

\section{Electric auxtethers enable spinrate control}

In E-sail plasma physics, a tether produces thrust per unit length which
is approximately proportional to the flow velocity of the plasma (equation 3 of \citet{RSIpaper}):
\begin{equation}
\frac{dF}{dz} = 0.18 \max\left(0,V_0-V_1\right) \sqrt{\epsilon_0
  P_\mathrm{dyn}^\perp}.
\label{eq:dFdz}
\end{equation}
Here $V_1=(1/2)m_p v_\perp^2/e \approx 1\,$kV is voltage corresponding
to solar wind proton kinetic energy, $V_0$ is the tether voltage and
$P_\mathrm{dyn}^\perp=\rho v_\perp^2$ is the solar wind dynamic
pressure expressed in terms of the solar wind mass density $\rho$ and
the solar wind tether-perpendicular velocity $\mathbf{v}_\perp$. More
accurate and more complicated thrust formulas also exist
\citep{RSIpaper}, but the assumption that the tether-parallel velocity
causes no propulsive effect remains exact as long as the tether is
much longer than the radius of the electron sheath that surrounds the
tether so that end effects can be ignored. This condition is typically
well valid since the tether length is of order 10-20 km while the
sheath radius at 1 au is $\sim 0.1$ km. In this section, the only
thing that we need from E-sail plasma physics is that a tether segment
generates a thrust vector which is aligned with the
segment-perpendicular component of the solar wind flow.

\begin{figure}[h]
\centering
\includegraphics[width=\columnwidth]{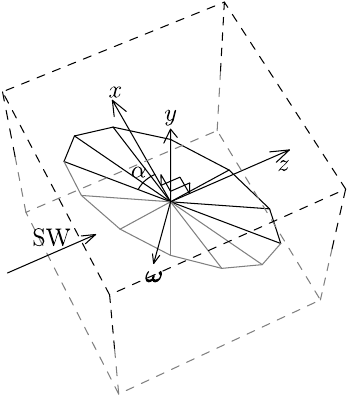}
\caption{
Three-dimensional schematic presentation of spinning planar E-sail
inclined at angle $\alpha$ with respect to SW flow ($\alpha$
lies in the $xz$ plane). Lines below $y=0$ plane are drawn in
greyscale to ease visualisation. The $z$ coordinate is along the SW.
}
\label{fig:ElAux3D}
\end{figure}

We consider an E-sail as in Fig.~\ref{fig:ElAux3D} where the auxiliary
tethers (auxtethers) are metallic and can be biased at high voltage,
similarly to the maintethers. A segment of an auxtether then generates
E-sail thrust which is perpendicular to it. Our aim is then to show
that if the auxtether voltages can be controlled independently from
the maintether voltages, spinrate control becomes possible.

\begin{figure*}[t]
\begin{center}
\includegraphics[width=\textwidth]{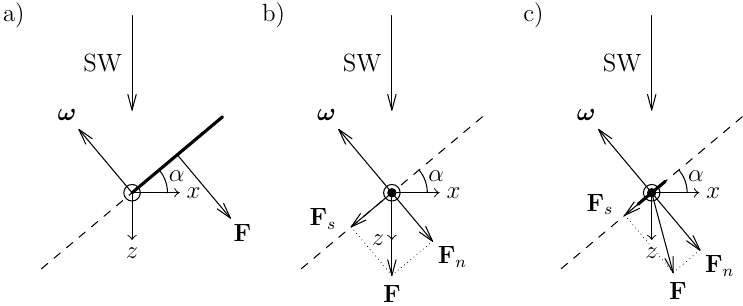}
\caption{
E-sail force components. (a) Maintether in $xz$ plane, (b) maintether
parallel to $y$, (c) maintether parallel to $y$ plus auxtether
segment.
}
\label{fig:ElAux2D}
\end{center}
\end{figure*}

Figure \ref{fig:ElAux2D}a again shows an E-sail inclined at angle $\alpha$
to the SW flow, but now viewed from the top, antiparallel to the $y$
axis. Consider a maintether in the $xz$ plane i.e.~in the plane of
Fig.~\ref{fig:ElAux2D}a. The maintether generates a thrust
vector $\mathbf{F}$ which is perpendicular to itself.

Figure \ref{fig:ElAux2D}b shows the same maintether $90^{\circ}$
rotation later when it is parallel to $y$ axis. Now, because the
tether is perpendicular to the SW, its thrust vector $\mathbf{F}$ is
aligned with the SW. We decompose $\mathbf{F}$ in spinplane component
$\mathbf{F}_s$ and spinplane normal component $\mathbf{F}_n$. The
spinplane component $\mathbf{F}_s$ brakes the tether's spinrate when
it moves upstream and accelerates it $180^{\circ}$ rotation later, and
the net effect vanishes. This means that by modulating maintether
voltages alone, one cannot change the sail's spinrate if one wants to
keep the sail's orientation constant. Modulation of maintether
voltages can tilt the sail which also changes the spinrate, but
independent control of the spinrate and orientation is not possible if
maintether modulation is the only available control. The secular
spinrate change effect arises because when orbiting the Sun, the Sun
moves with respect to the inertial frame (the celestial sphere defined
by distant stars) and the sail must track this motion. Doing so
requires application of torque because in the absence of torque the
angular momentum vector of the sail tends to be conserved i.e. the
spin axis tends to point to the same distant star. Tracking the Sun's
motion is equivalent to continuous turning of the sail, which changes
the spinrate as a byproduct if performed by modulating the maintether
voltages. The spinrate change occurs in this case because in order to
tilt the sail, the maintethers must be modulated unsymmetrically in
the $y$ direction so that symmetry in their upstream/downstream motion
is broken and a net spinrate change results. For an equivalent
explanation in the Sun-pointing orbital reference frame, see Figure 8
of \citep{paper14}.

Panel \ref{fig:ElAux2D}c is the same as panel \ref{fig:ElAux2D}b, but
we have added a charged auxtether segment at the tip of the
maintether.  The thrust vector $\mathbf{F}$ is now a vector sum of the
maintether thrust and the auxtether thrust. The maintether thrust is
still along the SW flow as it was in \ref{fig:ElAux2D}b, but the
auxtether's thrust contribution is perpendicular to the auxtether,
i.e.~perpendicular to the spin plane. As a result, $\mathbf{F}$ is not
aligned with the SW and the ratio $F_s/F_n$ depends on the ratio of
the auxtether thrust versus the maintether thrust. In particular, by
modulating the auxtether and maintether voltages separately, the ratio
$F_s/F_n$ can be different when the maintether is parallel or
antiparallel with the $y$ axis. By having the same $F_n$ but different
$F_s$ in the upstream and downstream portions of the maintether's
rotation cycle, we can modify the sail's spinrate while keeping its
orientation fixed. Separate control of sail spinrate and spinplane
orientation becomes possible because one has two control parameters in
each angular segment, namely maintether voltage and auxtether voltage.

\section{TI tether rig}
\label{sect:TI}

\begin{figure}[h]
\centering
\includegraphics[width=\columnwidth]{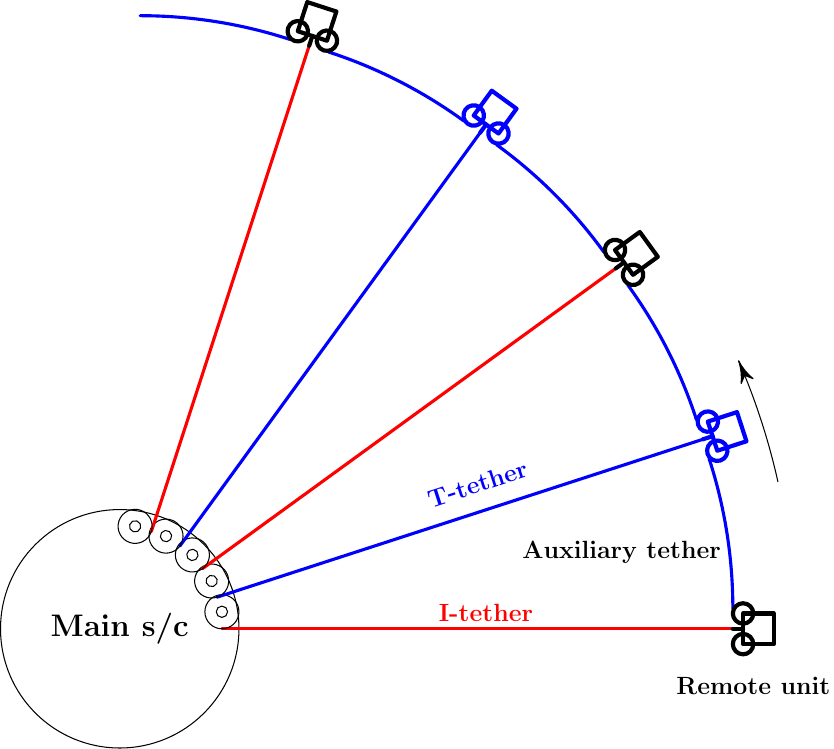}
\caption{
Schematic presentation of the TI tether rig.
}
\label{fig:schematic}
\end{figure}

To enable separate control of auxtether and maintether voltages, one
could use various technical means, for example, each remote unit could
carry a potentiometer or other means of regulating the auxtether
voltage between zero and the maintether voltage. However, we propose a
simpler arrangement where the remote units need no active parts. We
propose that even-numbered maintethers are such that their remote unit
is galvanically connected with both the left-side and right-side
auxtethers (Fig.~\ref{fig:schematic}, blue), while odd-numbered
maintethers are electrically insulated from the remote unit
(Fig.~\ref{fig:schematic}, red). We call the even-numbered tethers the
T-tethers because of the T-shaped shape of the blue equipotential
region, and odd-numbered tethers are correspondingly called I-tethers.

In a given angular sector of the sail, we can effectively increase
(decrease) the auxtether voltages by setting T-tethers to higher
(lower) voltage than I-tethers. The auxtethers are always at the same
potential as their associated T-tether so that no potentiometers or
other functional parts are needed on the remote units. Two types of
remote units are needed: ones that provide galvanic connection between
the maintether and the two auxtethers, and ones that provide an
insulating connection between all three connecting tethers. As usual,
the remote units contain reels of the auxtethers which are used during
deployment phase. During propulsive flight, no functionality is
required from the remote units. The units only have to continue to
provide the mechanical and electrical connection which is of galvanic
and insulating type of even and odd-numbered units,
respectively. Because of the presence of T-tethers and I-tethers, we
call the design as a whole the TI tether rig.

\section{Control algorithm}
\label{sect:controlalgorithm}

The control algorithm consists of six throttling factors which are
multiplied together at the end (Eq.~\ref{eq:f}) to yield the
time-dependent voltage throttling factor for each maintether. The six
factors and their qualitative roles are introduced in Table
\ref{tab:factors}.

\begin{table}[h]
\centering
\caption{The six throttling factors.}
\begin{tabular}{ll}
\hline
$f_1$ & Turning the spinplane \\
$f_2$ & Maintaining the spinplane \\
$f_3$ & Changing the spinrate \\
$f_4$ & Damping collective oscillations \\
$f_5$ & Damping oscillations of tethers \\
$f_6$ & Setting thrust to wanted value \\
\hline
\end{tabular}
\label{tab:factors}
\end{table}

Before defining the six throttling factors, we discuss some
preliminaries related to the general strategy of the control
algorithm. Let $\mathbf{r}=(x,y,z)$ be the remote unit's position
vector relative to the spacecraft and
$\hat{\mathbf{e}}_r=\mathbf{r}/r$ is the corresponding unit vector. We
denote the angular momentum of the tether rig by $\mathbf{L}$ and the
corresponding unit vector (spin axis vector) by
$\hat{\mathbf{s}}=\mathbf{L}/L$. The controller computes an
instantaneous angular momentum $\mathbf{L}_\mathrm{inst}$
approximately from imaged positions $\mathrm{r}$ of the remote units
and their velocities $\mathrm{v}$ found by finite differencing with
$\Delta t=2$ s timestep. The angular momentum $\mathbf{L}$ used by the
control algorithm below is a time-averaged version of
$\mathbf{L}_\mathrm{inst}$ which is obtained by continuously solving
the differential equation
\begin{equation}
\frac{d\mathbf{L}}{dt} = \frac{\mathbf{L}_\mathrm{inst}-\mathbf{L}}{\tau_\mathrm{L}}
\label{eq:L}
\end{equation}
where $\tau_\mathrm{L}=1200$ s is the timescale used in the
time-averaging.

We are now ready to give the detailed definitions of the six
throttling factors used by the control algorithm.

\subsection{Factor $f_1$}

The first throttling factor is
\begin{equation}
f_1 = \max\left[0,
1 - g_\mathrm{t} \hat{\mathbf{e}}_r \cdot
\left(\hat{\mathbf{s}} \times \hat{\mathbf{n}}_\mathrm{goal}\right)\right]
\label{eq:f1}
\end{equation}
where $g_\mathrm{t}=1.0$ is a greediness parameter for spinplane turning and
$\hat{\mathbf{n}}_\mathrm{goal}$ is the goal spin axis
orientation. The factor $f_1$ is responsible for turning the spinplane
when $\hat{\mathbf{s}}\ne\hat{\mathbf{n}}_\mathrm{goal}$. It modulates
the tether voltages so that the SW thrust applies a torque to
the tether rig.

\subsection{Factor $f_2$}

The second throttling factor $f_2$ takes care of keeping the spinplane
orientation constant. The second factor is
\begin{equation}
f_2 = (1-A)K + A
\label{eq:f2}
\end{equation}
where the 'spinplane keeper factor' $K$ is
\begin{equation}
K = \frac{1}{\vert \hat{\mathbf{n}}_\mathrm{SW} - \hat{\mathbf{e}}_r
(\hat{\mathbf{e}}_r \cdot \hat{\mathbf{n}}_\mathrm{SW}) \vert^2}
\label{eq:K}
\end{equation}
and the auxiliary factor
\begin{equation}
A = \frac{1}{1+N_\mathrm{w}/(2\pi)}.
\label{eq:A}
\end{equation}
The algorithm works moderately well also with $A=0$, but by
  numerical experimentation we found that it works better
if $A$ is computed from Eq.~(\ref{eq:A}). The denominator of $K$ is the
tether-perpendicular component of $\hat{\mathbf{n}}_\mathrm{SW}$. If
the tethers spin rapidly so that they move nearly in a plane without
coning, $K$ does not depend on tether phase angle. However, in a real
sail some coning occurs. Then the $K$ factor decreases and increases
thrust on the upwind and downwind orientations of the spinning tether,
respectively, to keep the total torque zero.

\subsection{Factor $f_3$}

The third throttling factor $f_3$ takes care of increasing or
decreasing the spinrate. First we define the spinrate increase factor $S$ by
\begin{equation}
S = g_\mathrm{s}\left[s_\mathrm{goal}-\frac{\vert\mathbf{L}\vert}{\vert\mathbf{L}(0)\vert}\right].
\end{equation}
Here $g_\mathrm{s}=2.0$ is the spinrate increase greediness factor and
$s_\mathrm{goal}$ is the goal for the relative spinrate, i.e.~the
angular mometum magnitude relative to the initial angular momentum
magnitude $\vert\mathbf{L}(0)\vert$. The throttling factor is given by
\begin{equation}
f_3 =
1 - \mathrm{clamp}\,\left(
\pm S \hat{\mathbf{v}}\cdot\hat{\mathbf{n}}_\mathrm{SW},
-c_\mathrm{st},c_\mathrm{st}\right).
\end{equation}
Here $\mathbf{v}$ is the instantaneous velocity of the remote unit
(relative to the spacecraft, similarly to $\mathrm{r}$) and
$c_\mathrm{st}=0.2$ is the maximum allowed amplitude of our sawtooth
tether modulation. Plus sign is selected for T-tethers and minus sign
for I-tethers. The function $\mathrm{clamp}$ forces the first argument
within given limits $a$ and $b$, $a\le b$. For any $x$,
$\mathrm{clamp}\,(x)$ is defined by
\begin{equation}
\mathrm{clamp}\,(x,a,b) = \max(a,\min(x,b))
\label{eq:clamp}
\end{equation}

The controller algorithm as described up to now works, but it does not
damp tether oscillations that are produced by SW variations
and the spinplane manoeuvres. Neither does it set the E-sail thrust to
a wanted value. The purpose of the remaining factors $f_4$, $f_5$ and
$f_6$ is to take care of these.

\subsection{Factor $f_4$}

For the first damping related factor, $f_4$, we measure the spin-axis aligned speed $v_\mathrm{s}$
(sign convention: positive sunward) of the remote units relative to
the spacecraft, averaged over the remote units. The measurement is
done by finite differencing the imaged remote unit angular positions
and the throttling factor is
\begin{equation}
f_4 = 1 + \min\left(0,g_\mathrm{d} \frac{v_\mathrm{s}}{v_\mathrm{tot}}\right)
\label{eq:f4}
\end{equation}
where $g_\mathrm{d}=3.0$ is greediness factor for damping and $v_\mathrm{tot}$
is the average rotation speed of the remote units with respect to the
spacecraft. The idea is that if the tether rig oscillates
collectively along the spin axis so that the tether cone angle changes
periodically, the oscillation is damped if voltages are slightly
throttled down when the rig is moving in the direction of the SW.

\subsection{Factor $f_5$}

The factor $f_4$ reduces collective oscillation of the whole tether
rig, but each tether can also oscillate individually like a guitar string between
the spacecraft and the remote unit. For reducing these a bit faster oscillations we
introduce throttling factor $f_5$. We measure the instantaneous
thrust force $\mathbf{F}_\mathrm{sc}$ acting on the spacecraft body (at 20 s
resolution) by an onboard vector accelerometer. Notice that $\mathbf{F}_\mathrm{sc}$ is the
force exerted on the spacecraft by the tethers which is usually not equal to the
total E-sail force exerted on the whole tether rig, except as an average
over a long enough time period. When $\vert \mathbf{F}_\mathrm{sc} \vert$
increases significantly, we apply overall throttling $f_5$ to tether
voltages where
\begin{equation}
f_5 = 1 - \mathrm{clamp}\,\left(
\tau_\mathrm{d5} \frac{1}{F_0}\frac{d\vert\mathbf{F}_\mathrm{sc}\vert}{dt},
0,d_\mathrm{max}.
\right)
\label{eq:f5}
\end{equation}
Here $\tau_\mathrm{d5}=1200$ s is a damping timescale parameter,
$d_\mathrm{max}=0.05$ is the maximum applied thrust reduction due to
damping and $F_0$ is the typical tether tension multiplied by the
number of tethers $N_\mathrm{w}$. We set the typical tension equal to the
tether tension in the initial state.

\subsection{Factor $f_6$}

The final throttling factor $f_6$ is used to settle the E-sail thrust
to a wanted value $F_\mathrm{goal}$. We estimate the E-sail thrust on
the tether rig by using the inertial coordinate frame equation
\begin{equation}
\mathbf{F}_\mathrm{rig} = \frac{d\mathbf{p}}{dt} +
\frac{m_\mathrm{rig}}{m_\mathrm{sc}} \mathbf{F}_\mathrm{sc}
\label{eq:Frig}
\end{equation}
where $\mathbf{p}$ is the momentum of the tether rig relative to the
spacecraft (determined by imaging and finite differencing the remote
unit angular positions), $m_\mathrm{rig}$ is the mass of the tether
rig and $m_\mathrm{sc}$ is the mass of the spacecraft body. The first term is due
to acceleration of the tether rig with respect to the spacecraft body
and the second term is due to acceleration of the spacecraft with
respect to an inertial frame of reference. The time average of the
first term is obviously zero, but its instantaneous value is usually
nonzero and it carries information about tether rig oscillations that
we want to damp. The instantaneous thrust exerted on the
whole system (spacecraft plus tether rig) is
\begin{equation}
\mathbf{F}_\mathrm{tot} =
\mathbf{F}_\mathrm{sc} + \mathbf{F}_\mathrm{rig}.
\label{eq:Ftot}
\end{equation}
From the instantaneous $\mathbf{F}_\mathrm{tot}$ we calculate a
time-averaged version $\mathbf{F}_\mathrm{tot}^\mathrm{ave}$ by
keeping on solving the time-dependent differential equation
\begin{equation}
\frac{d\mathbf{F}_\mathrm{tot}^\mathrm{ave}}{dt}
= \frac{\mathbf{F}_\mathrm{tot} -
  \mathbf{F}_\mathrm{tot}^\mathrm{ave}}
{\tau_\mathrm{d6}}
\label{eq:Ftotave}
\end{equation}
where $\tau_\mathrm{d6}=1200$ s is another damping timescale
parameter. Finally the overall throttling factor $f_6$ is calculated
as
\begin{equation}
f_6 = \mathrm{clamp}\,\left(
f_6^\mathrm{old}
+ \frac{\Delta t_\mathrm{d}}{\tau_\mathrm{d6}}
\frac{F_\mathrm{goal}-\vert \mathbf{F}_\mathrm{tot}^\mathrm{ave}
  \vert}
{F_\mathrm{goal}},
0,f_6^\mathrm{max}
\right)
\label{eq:f6}
\end{equation}
where $\Delta t_\mathrm{d}=20$ s is the timestep how often the damping
algorithm is called, $f_6^\mathrm{old}$ is the previous value of $f_6$
and $f_6^\mathrm{max}=1.01$ is $f_6$'s maximum allowed value. Equation
(\ref{eq:f6}) resembles solving a differential equation similar to
(\ref{eq:L}) and (\ref{eq:Ftotave}), except that (\ref{eq:f6}) also
clamps the solution if it goes outside bounds $(0,f_6^\mathrm{max})$.

\subsection{Combining the throttling factors}

The total throttling factor is 
\begin{equation}
f = \frac{f_1 f_2 f_3}{\max(f_1 f_2 f_3)} f_4 f_5 \min(1,f_6).
\label{eq:f}
\end{equation}
where the maximum is taken over the maintethers.

Factors $f_4$, $f_5$ and $f_6$ are updated at $\Delta t_\mathrm{d}=20$
s intervals while $f_1$, $f_2$ and $f_3$ are updated with $\Delta t=2$
s time resolution. The motivation for using slower updating of $f_4$,
$f_5$ and $f_6$ is only to save onboard computing power. The computing
power requirement is low in any case, but as a matter of principle we
want to avoid unnecessary onboard computing cycles.

Factors $f_4$ and $f_5$ make only small modifications to the total
throttling factor $f$. Despite this, their ability to damp tether rig
oscillations is profound.

The tether voltages are modulated by $f$. We assume in this paper that
the E-sail force depends linearly on $V$ so that we can achieve the
wanted force throttling by simply modulating the voltages by $f$. This
should be a rather good approximation (see equation 3 of
\citet{RSIpaper}). Were this assumption not made, the nonlinear
relationship, if any, should be modelled or determined experimentally and then
used during flight to map thrust modulation values $f$ into voltage
modulation values. Doing so is straightforward if such
relationship is known. Hence there is no loss of generality in making a working
assumption of a linear relationship between voltage and thrust.

\section{Simulation model}
\label{sect:sim}

We use a dynamical simulator which was built for simulating dynamical
behaviour of the E-sail tether rig \citep{SimDoc1,SimDoc2}. The
simulator models the E-sail as a collection of point masses, rigid
bodies and interaction forces between them. Also external forces and
torques can be included. The core of the simulator solves the ordinary
differential equations corresponding to Newton's laws for the
collection the bodies. The solver is an eight order accurate adaptive
Runge-Kutta solver adapted from \citet{NumericalRecipes3}. The solver
provides in practice fully accurate discretisation in time. The only
essential approximation is replacing continuous tethers by chains of
point masses connected by interaction forces that model their
elasticity. The E-sail force (a more accurate version of
Eq.~\ref{eq:dFdz} taken from \citet{RSIpaper}) is included in the model.
Table \ref{tab:simparams} summarises the main parameters of the simulation
used in this paper.

\begin{table}[h]
\centering
\caption{Simulation parameters.}
\begin{tabular}{ll}
\hline
Number of tethers $N_\mathrm{w}$ & 20 \\
Tether length & 10 km \\
Thrust goal $F_\mathrm{goal}$ & 100 mN \\
Solar distance & 1 au \\
Baseline tether voltage & 20 kV \\
Maximum tether voltage & 40 kV \\
Spacecraft body mass $m_\mathrm{sc}$ & 300 kg \\
Remote unit mass & 0.4 kg \\
Initial tether tension & 5 cN \\
Initial spin period & 2000 s \\
Tether linear mass density & $1.1\cdot 10^{-5}$ kg/m \\
Tether parallel wires & $3\times \phi{=}20\,\mu$m \\
Tether wire Young modulus & 100 GPa \\
Tether wire relative loss modulus & 0.03 \\
Remote unit imager resolution & $0.17^{\circ}$ \\
Onboard accelerometer noise & 1.5 $\mu g/\sqrt{\mathrm{Hz}}$ \\
Synthetic SW density & 7.3 cm$^{-3}$ \\
Synthetic SW speed & 400 km/s \\
Number of tether discr.~points & 10 \\
Placement of discretisation points & Parabolic \\
Number of auxtether discr.~points & 1 \\
Simulation length & 3 days \\
\hline
\end{tabular}
\label{tab:simparams}
\end{table}

The core of the simulator coded in C++ for high performance, while the
definition of the model (the collection of point masses, rigid bodies,
their interaction forces and external forces and torques) is coded in
Lua scripting language. One Lua function implements the control
algorithm described in Section \ref{sect:controlalgorithm} above. The
control algorithm needs only two types of sensors. Firstly, we need
imaging sensors to detect the angular positions of the remote units
with moderate angular $0.17^{\circ}$ resolution and 2 s temporal
resolution. The angular resolution requirement corresponds to about
2200$\times$530 pixels, either in a single panoramic imager or several
small imagers along the spacecraft's perimeter. Secondly, we need a
vector accelerometer onboard the main spacecraft, for which we assume
noise level of 1.5 $\mu g/\sqrt{\mathrm{Hz}}$. A low-noise low-noise
accelerometer such as Colibrys SF-1500 has noise level five times
smaller than this. The imager resolution and accelerometer noise level
were found by numerical experimentation. The chosen values are optimal
in the sense that smaller measurement error in sensors would not
noticeably improve the fidelity of the control and its oscillation
damping properties.

In Table \ref{tab:controlparams} we summarise the parameters of the
control algorithm, including its virtual sensors.

\begin{table}[h]
\centering
\caption{Default parameters of the control algorithm and its virtual sensors.}
\begin{tabular}{lll}
\hline
$d_\mathrm{max}$ & Maximum thrust reduction for $f_4$ & 0.05 \\
$f_6^\mathrm{max}$ & Maximum allowed $f_6$ & 1.01 \\
$F_\mathrm{goal}$ & Goal E-sail thrust & 100 mN \\
$g_\mathrm{d}$ & Greediness for damping in $f_4$ & 3.0 \\
$g_\mathrm{s}$ & Greediness for spinrate change & 2.0 \\
$g_\mathrm{t}$ & Greediness for spinplane turning & 1.0 \\
$\Delta t$ & Controller call interval & 2 s\\
$\Delta t_\mathrm{d}$ & Damper call interval & 20 s\\
$\tau_\mathrm{d5}$ & Timescale for damping oscillations & 1200 s\\
$\tau_\mathrm{d6}$ & Timescale for regulating thrust & 1200 s\\
$\tau_\mathrm{L}$ & Ang.~momentum averaging time & 1200 s\\
\hline
\end{tabular}
\label{tab:controlparams}
\end{table}

\section{Simulation results}
\label{sect:results}

All simulations start from an initial state where the sail rotates
perpendicular to the SW. Synthetic constant SW is used in first three
runs. In the last run, real SW is used. In all runs the thrust is
modulated by $1-\exp(-t/(4\mathrm{h}))$ so that it starts off
gradually from zero (a smooth transition from zero to one in a 4-hour
timescale). This is done to avoid inducing tether oscillations as an
initial transient: although the algorithm can damp such oscillations,
damping would not occur immediately.

\begin{figure}[h]
\centering
\includegraphics[width=0.82\columnwidth]{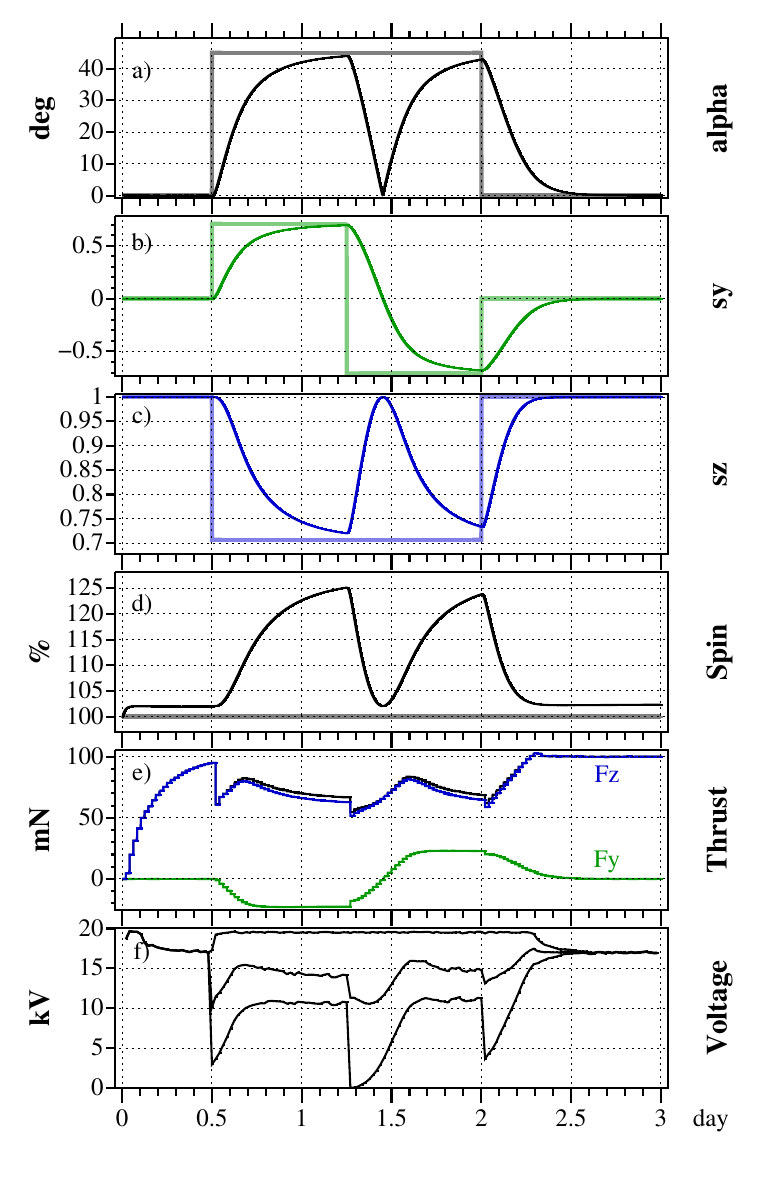}
\caption{
Result of Run 1. (a) angle $\alpha$ between SW
and spin axis; (b) $\hat{\mathbf{s}}_y$ ($y$ component of spin axis unit
vector $\hat{\mathbf{s}}$); (c) $\hat{\mathbf{s}}_z$ ($z$ component
of $\hat{\mathbf{s}}$); (d) spin angular momentum relative to initial angular
momentum in percent; (e) thrust along SW (blue, $F_z$), perpendicular
to it (green, $F_y$) and total (black); (f) tether instantaneous minimum, mean and maximum
voltages. In a-d, thicker grey and pastel lines show the commanded goal of each
parameter.
}
\label{fig:run1}
\end{figure}

In Run 1 (Fig.~\ref{fig:run1}), the tilt angle goal (panel a) is zero
until 0.5 days, then it is set to 45$^{\circ}$ where it remains for
1.5 days. The sail starts turning when the angle is set and reaches
almost $45^{\circ}$ angle after 0.75 days. Then the $\phi$ angle goal
(the polar angle of the spin vector) is
changed from 90$^{\circ}$ to -90$^{\circ}$ so that the sail starts turning
again, via zero to the opposite direction. At 2 days the $\alpha$ angle
goal is returned back to zero. Thus, Run 1 exercises a back and forth
swing of the tether rig. Spinrate regulation greediness parameter
$g_\mathrm{s}$ is set to zero in Run 1 so that we can observe the natural
tendency of the spinrate to vary during the turning manoeuvre. The
spinrate (Fig.~\ref{fig:run1}, panel d) increases up to 25\,\% from
the initial value when the sail reaches $\approx 45^{\circ}$ angle. The
increase is due to conservation of the sun-directed angular momentum
component $L_z$: $\vert \mathbf{L}\vert=\sqrt{L_x^2+L_y^2+L_z^2}$ must
increase if $L_x^2+L_y^2$ increases while $L_z$ remains constant.

The thrust direction (Fig.~\ref{fig:run1}, panel e) varies according
to the spinplane orientation. The total thrust is somewhat smaller
when the spinplane is actively turned, which is due to the fact some
tethers are then throttled in voltage (Fig.~\ref{fig:run1}, panel f).

\begin{figure}[h]
\centering
\includegraphics[width=0.85\columnwidth]{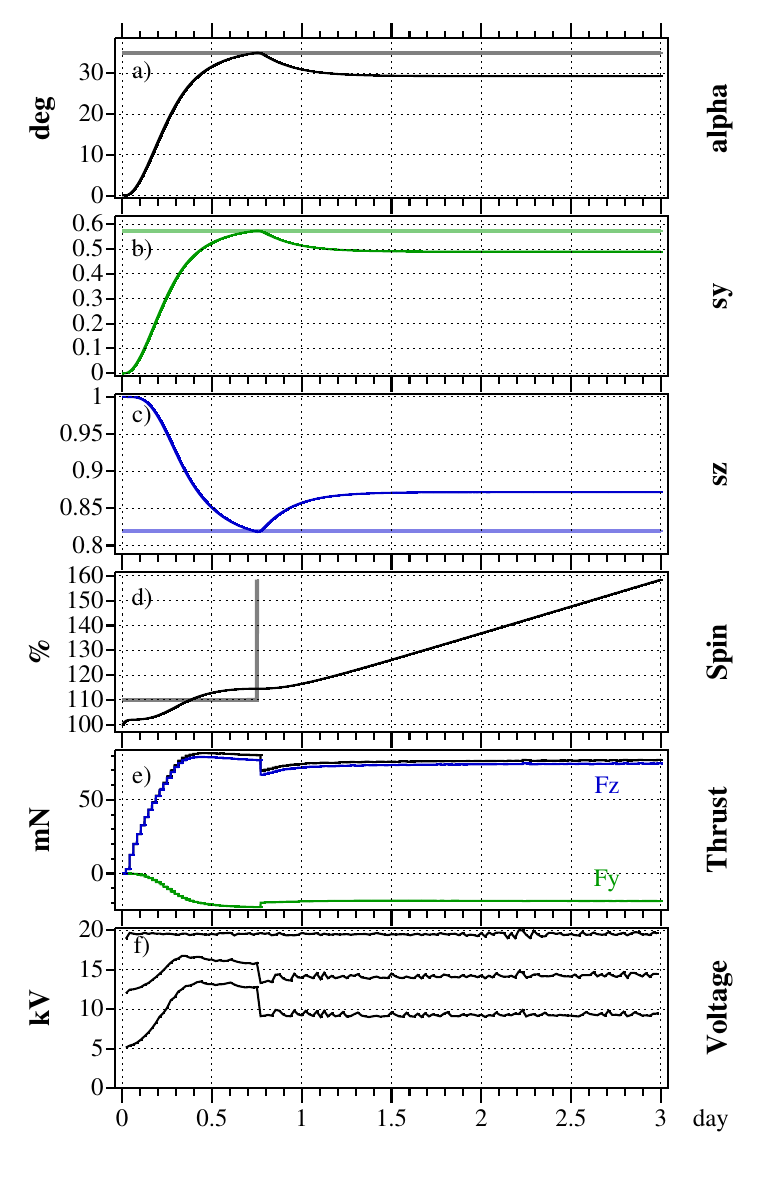}
\caption{
Same as Fig.~\ref{fig:run1} but for Run 2: demonstration of rapid spin increase.
}
\label{fig:run2}
\end{figure}

In Run 2 (Fig.~\ref{fig:run2}), the goal $\alpha$ angle is put to
$35^{\circ}$ throughout. The spinrate control greediness parameter $g_\mathrm{d}$
is put to its normal value of 2.0. The spinrate goal is 110\,\% spin
for the first 0.75 days and is put to very large value after that. The
controller turns the spinplane smoothly to $35^{\circ}$ which also
increases the spinrate moderately because of $L_z$ conservation. When
the spinrate goal is put high, the spinrate starts to increase almost
linearly, reaching 60\,\% increase at the end of the run which is 2.25
days since setting the spinrate goal high. As a byproduct of the
spinrate increase part of the algorithm, the sail angle
(Fig.~\ref{fig:run2}, panel a) decreases slightly from $35^{\circ}$ to
about $30^{\circ}$. The reason is that the spinrate modification and
tilt angle modification parts of the controller algorithm slightly
compete with each other because both use the same tether voltages for
actuation. We do not expect this competition to be a practical issue
because usually (to compensate the secular trend) the wanted spinrate
change is much slower than in Run 2. In any case, Run 2 shows that if
needed for any reason, the spinrate can be increased in a matter of
few days with the model sail.

\begin{figure}[h]
\centering
\includegraphics[width=0.85\columnwidth]{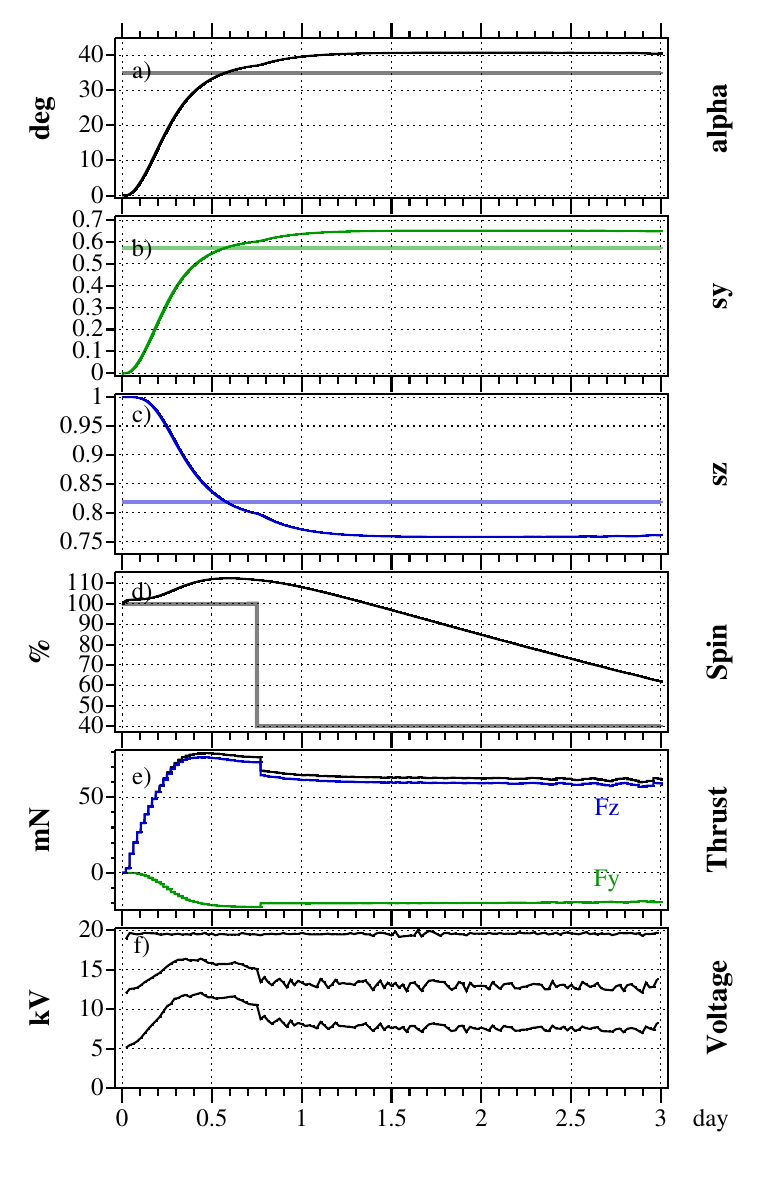}
\caption{
Same as Fig.~\ref{fig:run1} but for Run 3: demonstration of spin decrease.
}
\label{fig:run3}
\end{figure}

Run 3 (Fig.~\ref{fig:run3}) is similar to Run 2, but now we
demonstrate decreasing rather than increaseing of the spinrate. The
spinrate goal is put to 40\,\% at 0.75 days. The spin slows down
obediently. In this case the sail angle increases somewhat above the
goal value $35^{\circ}$.

\begin{figure}[h]
\centering
\includegraphics[width=0.85\columnwidth]{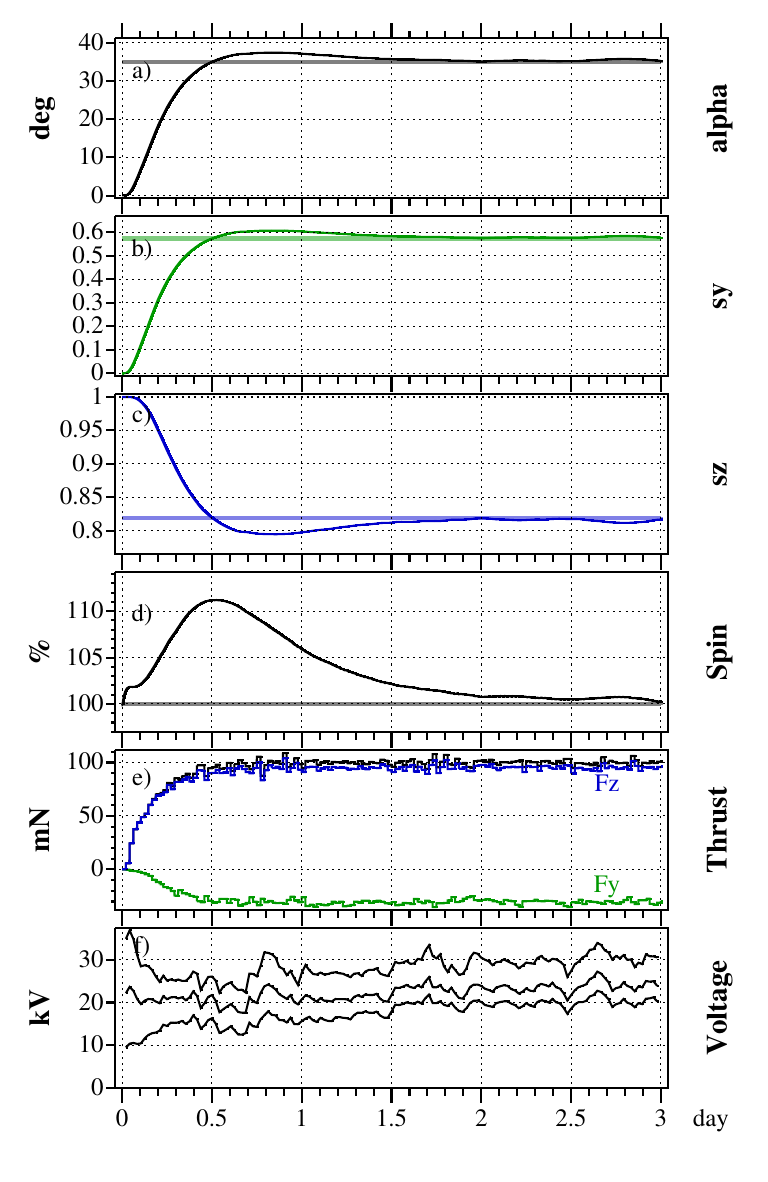}
\caption{
Same as Fig.~\ref{fig:run1} but for Run 4: typical use case of E-sail
with real SW.
}
\label{fig:run4}
\end{figure}

\begin{figure}[h]
\centering
\includegraphics[width=\columnwidth]{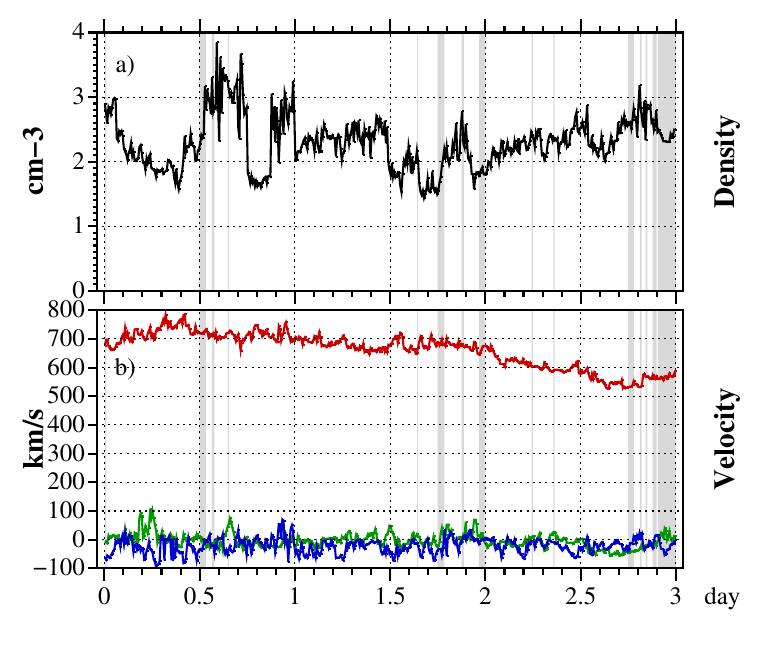}
\caption{
SW data used in Run 4 (Fig.~\ref{fig:run4}). (a)
plasma density, (b) SW velocity components (blue $x$,
green $y$, red $z$). Filled data gaps are shown as grey.
}
\label{fig:SW}
\end{figure}

Finally, in Run 4 (Fig.~\ref{fig:run4}) we simulate a typical use case of the E-sail. We set
the sail angle $\alpha$ goal to $35^{\circ}$ and the spinrate goal at
100\,\%. In Run 4 we also use real SW data to drive the E-sail
where $t=0$ corresponds to epoch January 1, 2000, 00:00 UT. The used
SW data comes from NASA/GSFCV's OMNI 1-minute resolution
dataset through OMNIWeb (Fig.~\ref{fig:SW},\citep{KingAndPapitashvili2005}).

\begin{figure}[h]
\centering
\includegraphics[width=\columnwidth]{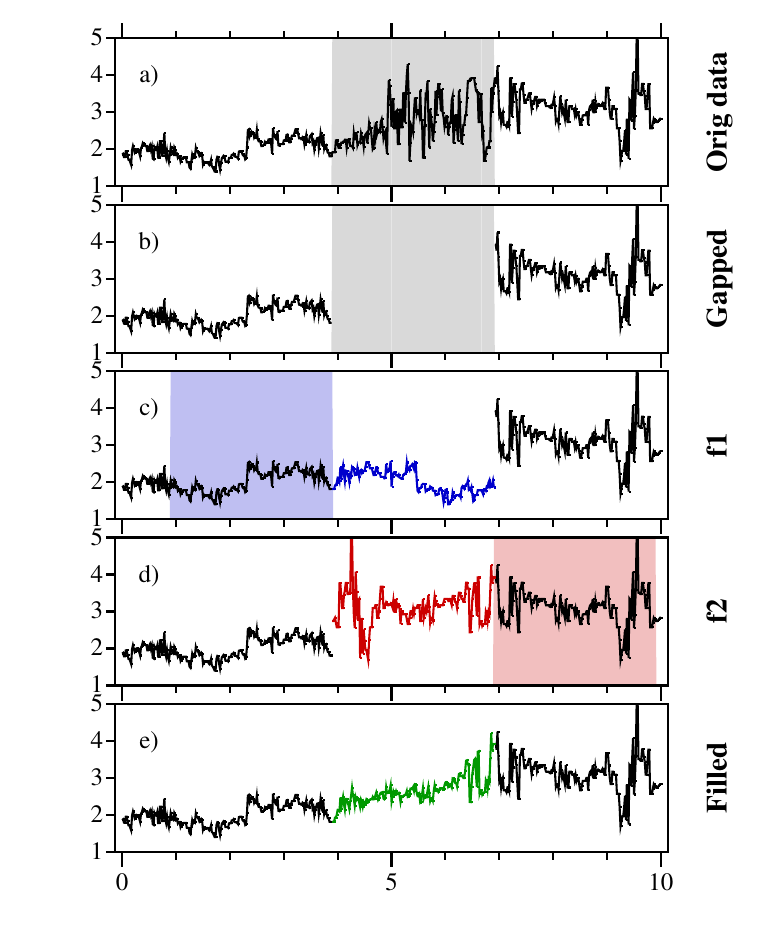}
\caption{
SW data gap filling algorithm. (a) original data, (b) original
data with gap removed, (c) gap filled by mirroring left side function,
(d) gap filled by mirroring right side function, (e) linear
interpolation of c and d removes jumps at gap boundaries. The data
shown in all panels is the solar wind plasma density in units of $\mathrm{cm}^{-3}$.
}
\label{fig:gaprm}
\end{figure}

The OMNI dataset contains data gaps, which we filled by the following
simple algorithm (Fig.~\ref{fig:gaprm}). Let $f(t)$ be the data which
has a gap at $t_1{<}t{<}t_2$.  Mirror the data before $t_1$ to make a
function $f_1(t)=f(2t_1-t)$. Now, function $f_1(t)$ fills the gap
$[t_1,t_2]$ with data that has the same spectral content as the real
data $f(t)\vert t{<}t_1$. The filler $f_1(t)$ has, however, a
discontinuity where the gap ends at $t_2$ and we return to real data
$f(t)\vert t{>}t_2$. To remedy this, we carry out a similar procedure
at the other end, mirroring data around $t_2$ to get
$f_2(t)=f(2t_2-t)$. Finally we construct the filler $\tilde{f}(t)$,
$t_1{<}t{<}t_2$, by linear interpolation between $f_1(t)$ and
$f_2(t)$: $\tilde{f}(t)=(1-u)f_1(t)+u f_2(t)$ where
$u=(t-t_1)/(t_2-t_1)$. The result is a gap-free solar wind time series
that has no discontinuous jumps and that retains as much as possible
the spectral properties of the true data.

Run 4 demonstrates numerically that the control algorithm correctly
tilts the sail to the wanted tilt angle and keeps it there, despite
variations of the solar wind. Tilting the sail causes the spinrate to
increase initially by $\sim 10$\,\% because of angular momentum
conservation, but the control algorithm later settles it back to the
commanded value. The algorithm accomplishes these tasks by using only
the two types of simulated sensors (with realistic noise components)
described in Section \ref{sect:sim}.

\section{Summary and conclusions}

We have presented a new E-sail design and its accompanying control
algorithm and sensor set which satisfies the following requirements:
\begin{enumerate}
\item Control of tether voltages from the main spacecraft is the only actuation mechanism.
\item Capability to control the orientation of the spin plane and
  thereby the orientation of the E-sail thrust vector.
\item Delivery of the wanted amount of E-sail thrust.
\item Spinrate acceleration and deceleration capability. With typical parameters, the
  spinrate modification control authority is many times larger than
  what is needed to overcome the heliocentric orbit Coriolis effect.
\item Remote units have no functionality requirements after deployment.
\item Both maintethers and auxtethers are biased and thereby
  propulsive.
\item Only two sensors are needed: remote unit angular position
  detection by imaging and accelerometer.
\item Moderate resolution sufficies for the imaging sensors.
\item The accelerometer should have low noise ($<1.5 \mu
  g/\sqrt{\mathrm{Hz}}$), but devices exist (e.g.~Colibrys SF-1500)
  whose noise level is even five times less.
\end{enumerate}

In the simulations of this paper we did not study deployment, but an
obvious question is if the spinrate increase capability of the
algorithm would be enough to deploy the sail in reasonable time. Based
on our preliminary analysis, the answer seems to be yes, provided that
deployment to a few hundred metre tether length is first achieved by
some other means.

Another future work that could be performed with our simulation is
systematic analysis of the average and maximum tether tension that
occurs during the run. Although not reported here, we have already
monitored tether tension in our simulations, and the version of the
control algorithm presented in this paper (Table
\ref{tab:controlparams}) was arrived at partly by trial and error
minimisation of the occurring maximum tether tension when thrust was
kept fixed. The peak tension is a measure of tether oscillations that
the control algorithm tries to keep at bay, hence low peak tension is
a figure of merit of the control algorithm. Typically the peak tension
can become some tens of percent higher than the average tension.

We think that the TI tether rig is a significant step forward in
E-sail design particularly because it enables full control of the
angular momentum vector while not requiring any functionality from the
remote units during flight. As a result, the secular spinrate problem
originally identified by \citet{paper14} gets solved in a simple way.

\section{Acknowledgement}



The work was partly supported by the European Space Agency. We
acknowledge use of NASA/GSFC's Space Physics Data Facility's OMNIWeb
service and OMNI data.








\end{document}